\documentclass[useAMS,usenatbib]{mn2e}
\usepackage{graphicx}
\usepackage{amsmath}
\usepackage{makecell}
\usepackage{macros}
\usepackage[usenames,dvipsnames,svgnames,table]{xcolor}
\usepackage{hyperref}
\definecolor{darkblue}{rgb}{0.0,0.0,0.3}
\hypersetup{colorlinks,%
            linkcolor=darkblue,urlcolor=darkblue,
            anchorcolor=darkblue,citecolor=darkblue}
\usepackage{txfonts}
\DeclareSymbolFont{cmletters}{OML}{cmm}{m}{it}
\DeclareMathSymbol{v}{\mathalpha}{cmletters}{"76}
\newcommand{\RedeclareMathOperator}[2]{\renewcommand{#1}{}\let#1\relax\DeclareMathOperator{#1}{#2}}

\newcommand\simless\lesssim
\newcommand\simgreat\gtrsim

\defcitealias{2015arXiv150802721B}{BT16}
\defcitealias{2014arXiv1408.3318P}{PK15}
\defcitealias{1997MNRAS.286..215K}{KA97}

\title[3D Kink Instability and FR Dichotomy]%
{Three-dimensional Relativistic MHD Simulations of Active Galactic \hbox{Nuclei Jets: Magnetic Kink Instability and Fanaroff-Riley Dichotomy}}
\author[Alexander Tchekhovskoy and Omer Bromberg]{Alexander Tchekhovskoy$^{1,
2}$\thanks{E-mail:
atchekho@berkeley.edu, omerb@astro.princeton.edu}\thanks{Einstein Fellow} and Omer Bromberg$^{3}$\thanks{Lyman Spitzer, Jr.\ Fellow}\\
$^{1}$Departments of Astronomy and Physics, Theoretical Astrophysics
Center, University of California Berkeley, Berkeley, CA 94720-3411\\
$^2$Lawrence Berkeley National Laboratory, 1 Cyclotron Rd, Berkeley,
CA 94720, USA\\
$^{3}$Department of Astrophysical Sciences, Peyton Hall, Princeton University, Princeton, NJ 08544, USA
}

\begin{document}

\date{Accepted. Received; in original form}
\pagerange{\pageref{firstpage}--\pageref{lastpage}} \pubyear{2015}

\maketitle

\label{firstpage}

\begin{abstract}
Energy deposition by active galactic nuclei jets into the ambient
medium can affect galaxy formation and evolution, the cooling of gas
flows at the centres of galaxy clusters, and the growth of the
supermassive black holes. However, the processes that couple jet
power to the ambient medium and determine jet morphology are poorly
understood. For instance, there is no agreement on the cause of the
well-known Fanaroff-Riley (FR) morphological dichotomy of jets, with
FRI jets being shorter and less stable than FRII jets.  We carry out
global 3D magnetohydrodynamic simulations of relativistic jets
propagating through the ambient medium.  We show that the flat
density profiles of galactic cores slow down and collimate the jets,
making them susceptible to the 3D magnetic kink 
instability. We obtain a critical power, which depends on the galaxy
core mass and radius, below which jets become kink-unstable within
the core, stall, and inflate cavities filled with
relativistically-hot plasma. Jets above the critical power stably escape
the core and form powerful backflows. Thus, the kink
instability controls the jet morphology and can lead to
the FR dichotomy.  The model-predicted dependence of the critical
power on the galaxy optical luminosity agrees well with
observations.
\end{abstract}


\section{Introduction}
\label{sec:introduction}

Active galactic nuclei (AGN) jet interaction with the
interstellar/intergalactic medium (ISM/IGM) is very rich and exhibits
different types of morphology of poorly understood origin. 
Some jets appear conical, show large-scale wiggles,
decelerate due to the interaction with the ambient medium
\citep{1993ASSL..103...95L,2002MNRAS.336.1161L}, %
like the M87 jet \citep*{1989ApJ...347..713H}. Their extended emission
at $1.4$~GHz
tends to peak less
than half-way from the nucleus to the outer edge; such jets are
classified as FRI jets
\citep{fr74}. FRII jets appear nearly straight, edge brightened, and end up in a hot
spot, like the Cygnus~A jet \citep{1984ApJ...285L..35P}. They are generally
more powerful than FRI jets (%
\citealt{owen_fri/il_1994,ledlow_20_1996}, with
recent revisions, \citealt{best_radio_2009}) and preferentially occur
in isolated, field galaxies \citep[see][and references
therein]{2015ApJ...805..101H}.  Although on average FRII jets are more
spatially extended than FRI jets, some FRI jets reach Mpc distances
\citep[][]{hardcastle_chandra_2002, laing_multifrequency_2008}.

There is no agreement on the physical reasons leading to
the FRI/FRII dichotomy.  These could include differences in the
central engine
\citep*[e.g.,][]{1995ApJ...451...88B,1999ApJ...522..753M} and
ambient medium \citep[e.g.,][]{2000A&A...363..507G}. Whereas the FRII
morphology was reproduced by pioneering
hydrodynamic simulations \citep*{clarke_numerical_1986}, the apparent
instabilities of FRI jets and their observed deceleration are much
more difficult to explain. Some of the possibilities include
Kelvin-Helmholtz (KH) instabilities in the shear layers %
(e.g.,
\citealt{1997MNRAS.286..215K}, hereafter
\citetalias{1997MNRAS.286..215K}; \citealt*{2005A&A...443..863P};
\citealt{2009ApJ...705.1594M}; \citealt{2010A&A...519A..41P}),
and mass entrainment from stellar winds
\citep{1994MNRAS.269..394K,2014MNRAS.441.1488P,wykes_internal_2015}.

 Magnetic fields are natural candidates to launch relativistic jets \citep*[e.g.][]{tch11}. Inferred to
be dynamically-important in the central regions
of many radio-loud AGN
\citep{2014Natur.510..126Z,2014Natur.515..376G,2015MNRAS.449..316N},
they may affect jet morphology by
suppressing KH mixing and initiating
current-driven instabilities, e.g., the 3D magnetic kink (``corkscrew'') instability
(\citealt*{nll07};
  \citealt{mignone_3d_jets_2010, 2012ApJ...757...16M}; 
  \citealt*{2014ApJ...781...48G,
  2014arXiv1408.3318P},  hereafter \citetalias{2014arXiv1408.3318P}; \citealt{2015arXiv150802721B}, hereafter \citetalias{2015arXiv150802721B}).
This
motivates a detailed 3D study of magnetic effects on the global
dynamics of jet-ISM/IGM interaction.
Extra care needs to
be taken in the way magnetic jets are initiated. In particular, different
values of the magnetic pitch, or the ratio of poloidal to toroidal
magnetic field, $B_p/B_\varphi$, at the injection point result in different jet
morphologies \citep{2014ApJ...781...48G}.  
To eliminate this uncertainty, we follow \citetalias{2015arXiv150802721B}
and set up our simulations to launch the jets the way nature does it: via the magnetised rotation
of a central object. 
We give numerical details in
Section~\ref{sec:prob}, present our results in Section~\ref{sec:instab}, and
conclude in Section~\ref{sec:concl}.

\begin{table}
\begin{center}
  \caption{Simulation setup parameters for the various models we present.}
 \begin{tabular}{@{}l l @{\quad} c @{\quad} c @{\quad} c @{$\;\;$} c @{$\;\;$} c @{$\quad$} c @{}} 
 \hline
 \thead{Model \\ name} & \thead{Resolution \\ ($N_r\times
   N_\theta\times N_\phi$)} & $\nu$ & $\Lambda_{\rm in}$ & \thead{$r_{\rm in}$} &
   \thead{$r_{\rm out}$} & \thead{$r_{\rm break}$} & $\displaystyle\frac{ct_{\rm sim}}{r_{\rm in}}$\\
\hline\hline
 P46    &  $256\times96\times192$ & $1$ & $9.3$ & $1$ & $10^5$        & --     & $2.1\times10^4$\\ 
 P46B2 &  $256\times96\times192$  & $1$ & $9.3$ & $1$ & $10^5$        & $10^2$ & $2.1\times10^4$\\ 
 P44    &  $256\times96\times192$ & $1$ & $4.3$ & $1$ & $10^5$        & --     & $5.4\times10^4$\\ 
 P44HR  &  $320\times96\times192$ &$1.4$& $4.3$ & $1$ & $10^3$        & --     & $3.4\times10^4$\\ 
 \hline
\end{tabular}\label{tab:models_grid}
\end{center}
\end{table}

\section{Numerical Method and Problem Setup}
\label{sec:prob}
We use global time-dependent relativistic 3D
magnetohydrodynamic (MHD) numerical simulations in order to study the
propagation of magnetised jets in an ambient medium characteristic of
AGN. We carry out the  simulations with the {\sc harm} code 
(\citealt*{gam03}; \citealt{nob06,tch07,mb09,tch11} and 
use modified spherical polar coordinates ($r$, $\theta$,
$\varphi$) described below that span the range $(r_{\rm in},r_{\rm out})\times(0,\pi)\times(0,2\pi)$.

\begin{figure}
  \centering
\includegraphics[width=0.9\columnwidth]{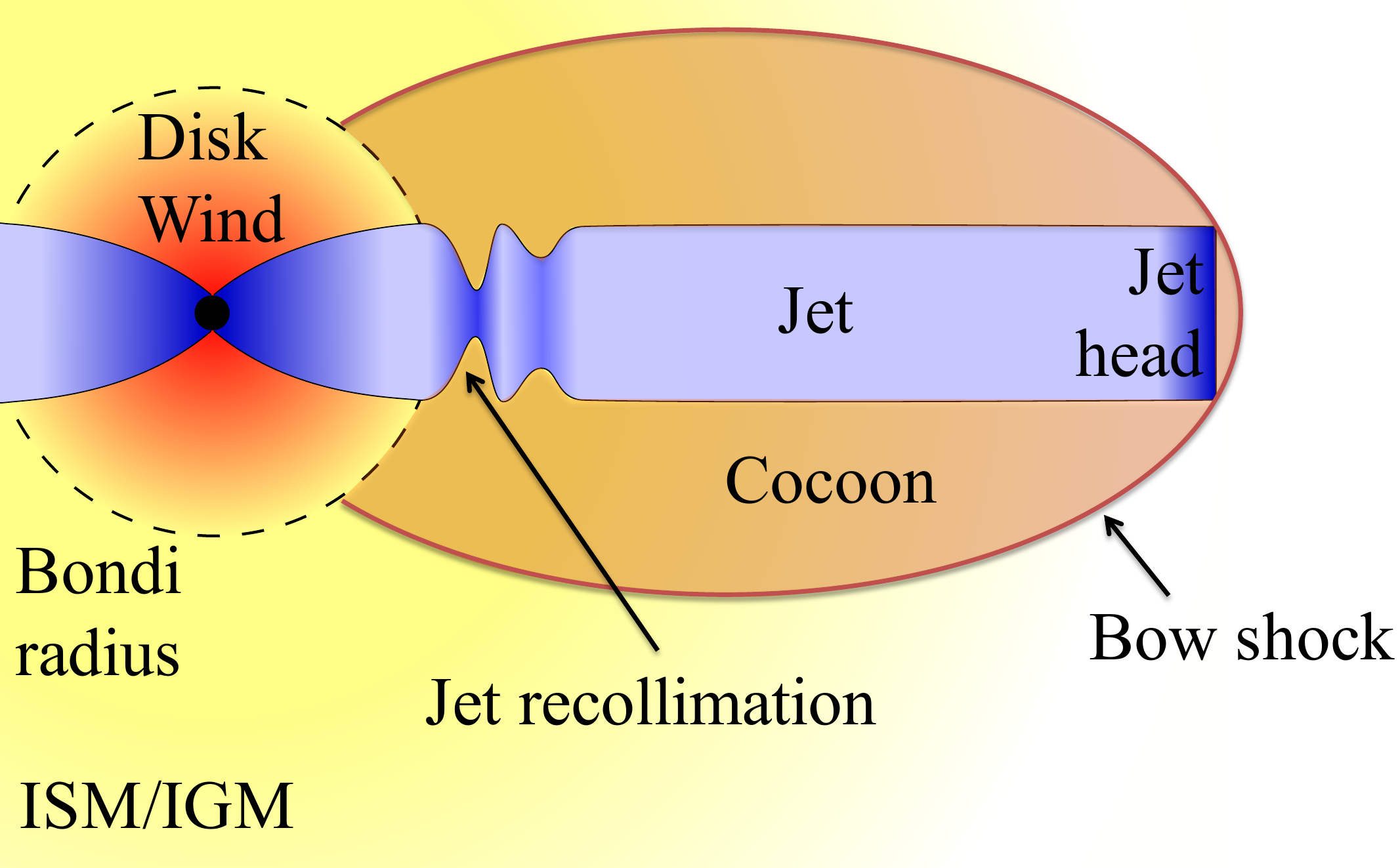}
\caption{Cartoon depiction of an AGN central engine. The jets (blue)
  collimate against the accretion disc wind (red) and assume a
  parabolic shape inside the Bondi radius (dashed line). Once outside,
  they start interacting with the ambient medium (ISM/IGM, shown in
  yellow).  As the jets adjust to this change in the ambient density
  profile, they go through a series of recollimation events and
  eventually settle into a near-cylindrical configuration. 
  At its head, the jet drills through the ISM/IGM and sends out a bow
  shock. The shocked ambient medium and the jet exhaust form a cocoon
  (brown) that collimates the jet outside the Bondi radius.
}
\label{fig:cartoon}
\end{figure}

Fig.~\ref{fig:cartoon} shows a cartoon of the AGN. Within the 
\emph{Bondi radius}, $r<r_{\rm B}$, the thermal pressure of the ISM/IGM
cannot support the gas against gravity. Here, black hole (BH) powered jets %
propagate unimpeded, are collimated by the accretion disc wind, %
and have a
parabola-like shape, as seen in the M87 galaxy
\citep{2013ApJ...775..118N} and the numerical simulations of
jet formation \citep*{mck06jf,hk06,tch11,tch12proc,tch12a,2015ASSL..414...45T}.  At
$r\gtrsim r_{\rm B}$, 
the jet shape
is observed to change from parabolic to conical, at least for the M87
galaxy  \citep{2013ApJ...775..118N}. It is plausible that around this
distance the jets
start to interact with the ISM/IGM,
causing them to undergo a series of recollimations
\citepalias{2015arXiv150802721B} that appear as a series of
stationary features, such as HST-1, seen in the M87 jet
(\citealt*{1999ApJ...520..621B}; \citealt{2013ApJ...774L..21M}).

Motivated by this, our fiducial choice for the position of the inner
radial boundary $r_{\rm in}$ is the Bondi radius, which we take to be
$r_{\rm B} = 0.1$~kpc \citep[see, e.g.,][]{2015MNRAS.451..588R}, and place the outer boundary at a large
distance, $r_{\rm out}\gg r_{\rm in}$, so that the transients do not
reach it in a simulation time (see Tab.~\ref{tab:models_grid}). We fill
the domain with a cold, spherically-symmetric density
distribution described in Sec.~\ref{sec:instab} and neglect gravity. 
We model
the inner boundary as a perfectly conducting magnetised sphere
threaded with a laterally-uniform radial magnetic flux.\footnote{To avoid the
interaction of the jets with the polar singularity, we orient the
rotational axis along the $x-$direction and collimate the radial grid
lines toward it in order to resolve the jets well, with the angle that
a radial grid line makes with the $x-$axis scaling as
$\chi\propto r^{-\nu/2}$ \citepalias[see][for
details]{2015arXiv150802721B}.}  At the beginning of the simulation, we
spin the polar caps of the sphere within $50^\circ$ of the rotational
axis at an angular frequency $\Omega=0.8c/r_{\rm in}$.  The initiation
of jets via the rotation at the base leads to a natural degree of magnetic
field azimuthal winding, a crucial factor that controls the stability
of magnetised jets \citepalias{2015arXiv150802721B}. The jets are
initially highly magnetised, with magnetisation at the inner radial
boundary, $\sigma \equiv 2p_{\rm m}/\rho c^2=25\gg1$, where $p_{\rm
m}$ is the magnetic pressure and $\rho$ is the fluid-frame mass
density in the jet.

\begin{table}
\begin{center}
  \caption{Simulations scaled
    to physical systems of different size and power. We indicate scaled
    versions of the models with suffixes.}
 \begin{tabular}{@{}l @{\;} l @{\quad} c @{\quad} c @{\quad} c @{\quad} c @{\quad} c @{\quad} c @{\quad} c @{}} 
 \hline
 \thead{Model \\ name} & \thead{Suffix} & $\Lambda_{\rm in}$ & \thead{$r_{\rm in}$\\{\rm [kpc]}} &
   \thead{$r_{\rm out}$\\{\rm [kpc]}} & \thead{$r_{\rm break}$\\{\rm [kpc]}} & \thead{$n_{\rm kpc}$\\${\rm [cm^{-3}]}$} & $\displaystyle\frac{L_j}{10^{45}}$ & \thead{$t_{\rm sim}$\\{\rm [Myr]}} \\
\hline\hline
 P46    &    & $9.3$ & $0.1$ & $10^4$        & --   & $0.2$ & $15$ & $6.7$\\ 
 P46B2  &    & $9.3$ & $0.1$ & $10^4$        & $10$ & $0.2$ & $15$ & $6.7$\\ 
 P44    &    & $4.3$ & $0.1$ & $10^4$        & --   & $0.2$ &$0.15$& $17$\\ 
 P44    & nx5 & $4.3$ & $0.1$ & $10^4$       & --   & $1.0$ &$0.75$& $17$\\ 
 P44    & rx5 & $4.3$ & $0.5$ & $5\times10^4$& --   & $0.2$ &$0.75$& $85$\\ 
 \hline
\end{tabular}\label{tab:models_phys}
\end{center}
\end{table}

\begin{figure}
  \centering
\includegraphics[width=\columnwidth]{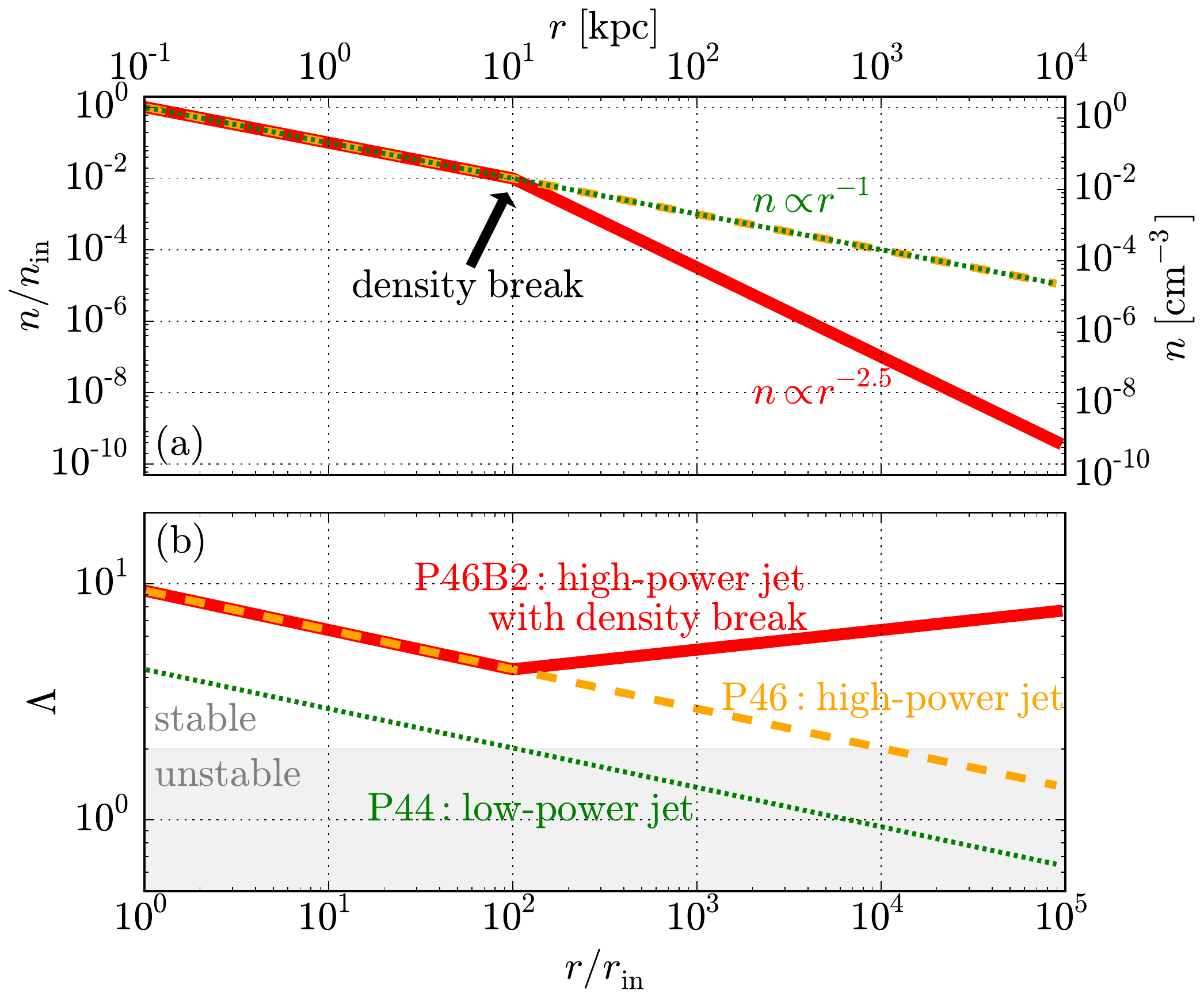}
\caption{{(panel a):} Ambient medium density profiles of our simulations. We consider
  flat 
  power-law density profiles at small distances ($n\propto r^{-1}$)
  that, in some models, steepen at larger radii ($n\propto
  r^{-2.5}$) to mimic the edge of the galaxy/cluster core. {(panel b):} Jet
  stability parameter versus radius in our simulations. The high-power
jets maintain stability out to Mpc distances, whereas the low-power
jets are expected to become unstable around $10$~kpc.}
\label{fig:lambda}
\end{figure}

\section{\hbox{Kink Mode and Jet Morphology}}
\label{sec:instab}
Of current-driven, 3D instabilities,
the most serious is the kink ($m=1$) mode. It causes the jets
to move bodily sideways and develop helical motions
\citep*{2000A&A...355..818A,nlt09}. To evaluate its potential to
disrupt the jets, \citetalias{2015arXiv150802721B} computed the ratio
of the instability growth timescale, evaluated as the time it takes an
Aflv\'en wave to travel around the jet $10$ times, to the time for a
fluid element to travel from the base to the tip of the
jet. %
This gives us the stability parameter,
\begin{equation}
  \label{eq:lambda}
  \Lambda = 2 \frac{\gamma_j\theta_j}{0.03} = K\times\left(\frac{L_j}{n m_p r^2 \gamma_j^2 c^3}\right)^{1/6},
\end{equation}
where, $n$ is the ambient medium
number density, $m_p$ is the proton mass, $r$ is the distance along
the jet, $K = 20(2\pi/9)^{1/2}[\pi(5-\alpha)(3-\alpha)/6]^{1/3}$ is
a constant prefactor,
and $\alpha = -{\rm d}\log n/{\rm d}\log r$ is the slope of the
ambient density profile \citepalias{2015arXiv150802721B}. Unless stated otherwise, we will assume that the jet plasma moves at a
mildly relativistic velocity, $\beta_j\equiv v_j/c\approx1$ and
$\gamma_j\equiv(1-\beta_j^2)^{-1/2}\approx1$, and is highly
magnetised, $\sigma\gg1$.
Equation~\eqref{eq:lambda} shows that
tightly collimated jets are the most susceptible to the kink instability:
in fact, if $\Lambda\lesssim\Lambda_{\rm crit}\equiv2$, or $\theta_j\lesssim
\theta_{\rm crit} \equiv 0.03/\gamma_j$, the
kink instability has sufficient time to develop and can
disrupt the jets and cause them to stall
\citepalias{2015arXiv150802721B}.

We consider two fiducial models: model P44,
representative of a weak FRI-like jet of power
$L_j\approx 1.5\times 10^{44}$ erg~s$^{-1}$, and model P46, 
representative of a powerful FRII-like jet of $L_j\approx 1.5\times 10^{46}$
erg~s$^{-1}$. Figure~\ref{fig:lambda}(a) shows the chosen initial ambient
density profile: a power-law,
$n=n_{\rm kpc} (r/{\rm kpc})^{-\alpha}$, with normalisation
$n_{\rm kpc} = 0.2$~cm$^{-3}$ and slope $\alpha = 1$ characteristic of
cores of elliptical galaxies such as M87 (see
Tab.~\ref{tab:models_grid} and
\citealt{1984ApJ...278..536S,2015MNRAS.451..588R}).  Since the jets in
powerful FRII sources often leave the host galaxy/cluster core, we consider
model P46B2 with a steeper density profile outside of the core,
$r>r_{\rm break}$, where the density is multiplied by a factor
$(r/r_{\rm break})^{-1.5}$.  We adopt $r_{\rm break}=10^2r_{\rm in}$,
leading to a sufficient scale separation between $r_{\rm in}$ and
$r_{\rm break}$ (see Fig.~\ref{fig:lambda} and
Tab.~\ref{tab:models_grid}).  By changing the length unit while
holding $\Lambda_{\rm in}=\Lambda(r_{\rm in})$ constant,
our simulations can be rescaled to different jet powers and ambient
densities (see models P44nx5 and P44rx5 in Tab.~\ref{tab:models_phys}).

As the jets run into the ISM/IGM, their dynamics is regulated by their
ability to drill through the ambient gas. In flat density profiles with
$\alpha < 2$, the mass per unit distance ahead of the jets increases
with distance. This causes the jets to slow down and collimate into
smaller opening angles, and become less stable, $\theta_j\propto\Lambda\propto
r^{-(2-\alpha)/6}$  (see eq.~\ref{eq:lambda} and Fig.~\ref{fig:lambda}b). Hence, in an $\alpha = 1$ environment, such as in
models P44 and P46, jets eventually become unstable, reaching
$\theta_j\lesssim\theta_{\rm crit}$ or equivalently
$\Lambda \lesssim 2$ (shaded area in Fig.~\ref{fig:lambda}b). This
occurs at a critical distance \citepalias{2015arXiv150802721B},
\begin{equation}
  \label{eq:rcrit}
  r_{\rm crit} = 7\;{\rm kpc} \times \left(\frac{L_j}{10^{44}\ {\rm erg\;s^{-1}}}
        \times\frac{0.2\ {\rm cm^{-3}}}{n_{\rm kpc}}\right)^{1/(2-\alpha)}.
\end{equation}
Fig.~\ref{fig:lambda}(b) shows that in the low-power model P44 the jets cross into the unstable region at
$r\gtrsim10$~kpc. In contrast, in the high-power model P46 the jets
maintain $\Lambda>2$ until Mpc-scales, suggesting that they remain
stable out to these large distances. In the model with the density
break, P46B2, the jets never cross into the instability region: as
they enter the steep density profile region ($\alpha>2$), they become
progressively more stable as they propagate away. To sum up, 
we expect low-power jets to become unstable and disrupt well within
the galaxy/cluster cores, and the high-power jets to remain stable
out to large distances, well outside the galaxy.

\begin{figure}
  \centering
\includegraphics[width=\columnwidth]{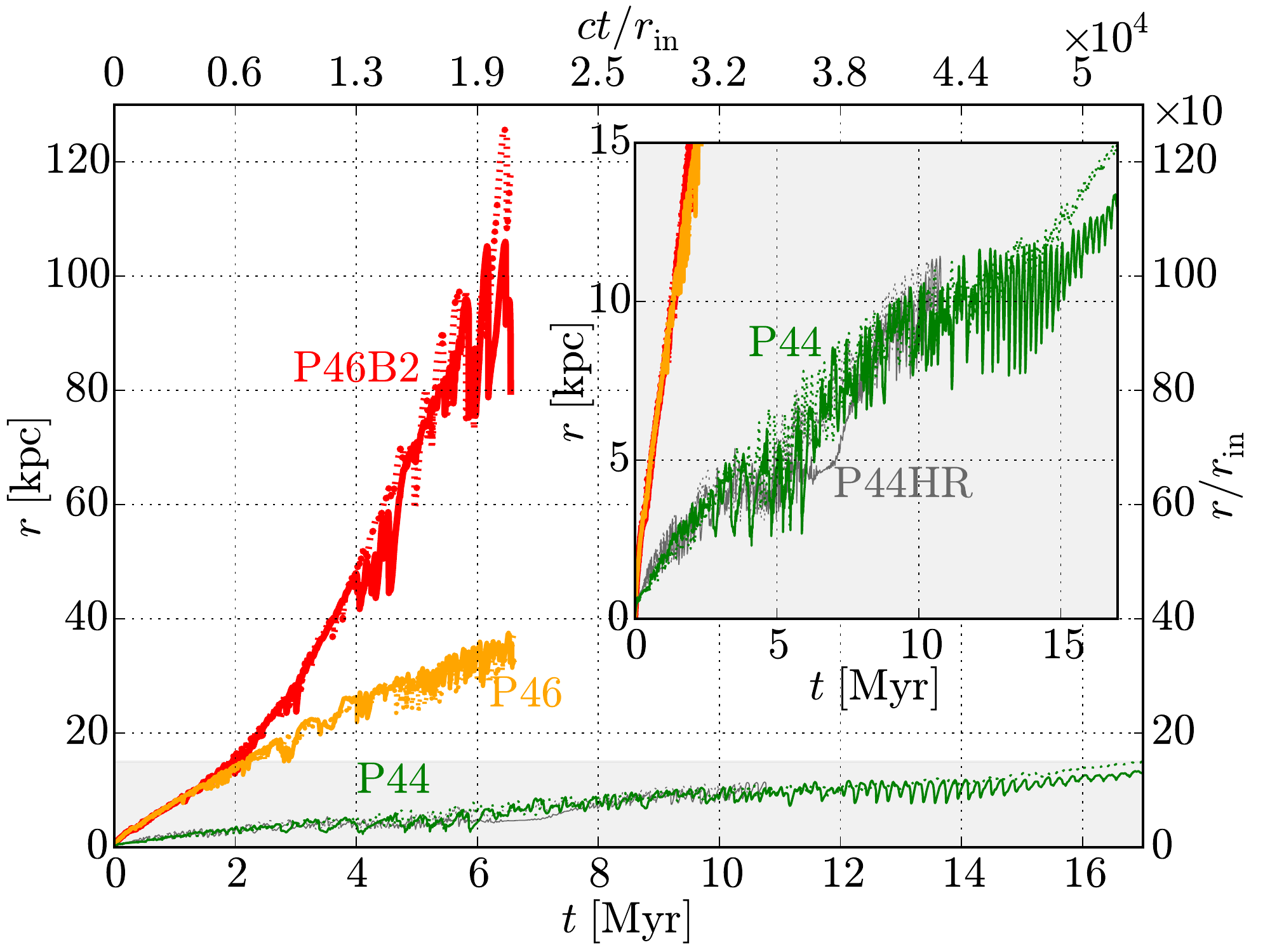}
\caption{Position of jet head versus time in different models
  considered in this work. Solid lines show the right
  and dotted lines left jet. The high-power jet with the density
  break (model P46B2, red lines),
  reaches $\sim100$~kpc and that without the density break (model P46, orange lines), reaches $\sim 30$~kpc after
  $6$~Myr. The inset shows with the green lines the zoom-in on the
  low-power jet model, P44, in which the jet propagates approximately
  $10$ times slower and stalls at $t\sim3{-}6$~Myr, with the
  head position showing large-amplitude oscillations around $r\sim
  5$~kpc. As a result, the jet inflates large
  cavities seen in Fig.~\ref{fig:dich}(c). Shown in grey, jet head
  position in the model P44HR, which has twice as high effective
  resolution in all $3$ dimensions, closely tracks that in the model
  P44; hence our models are well-resolved.}
\label{fig:zh_vs_t}
\end{figure}

Fig.~\ref{fig:zh_vs_t} quantifies jet stability to current-driven 3D
instabilities by showing the position of jet
head versus time for the first three models in
Tables~\ref{tab:models_grid} and \ref{tab:models_phys}. The jets in
the high-power model P46 decelerate, as expected in
flat density profiles with $\alpha < 2$
\citepalias{2015arXiv150802721B}.  Similarly, the jets in model P46B2
decelerate within the flat density core, $r< r_{\rm break}=10$~kpc,
but accelerate outside, as the density slope steepens to
$\alpha = 2.5$. As a result, at $6$~Myr the jets in model P46B2
reach $r\sim100$~kpc, three times larger than in model P46.

\begin{figure}
  \centering
\includegraphics[width=\columnwidth]{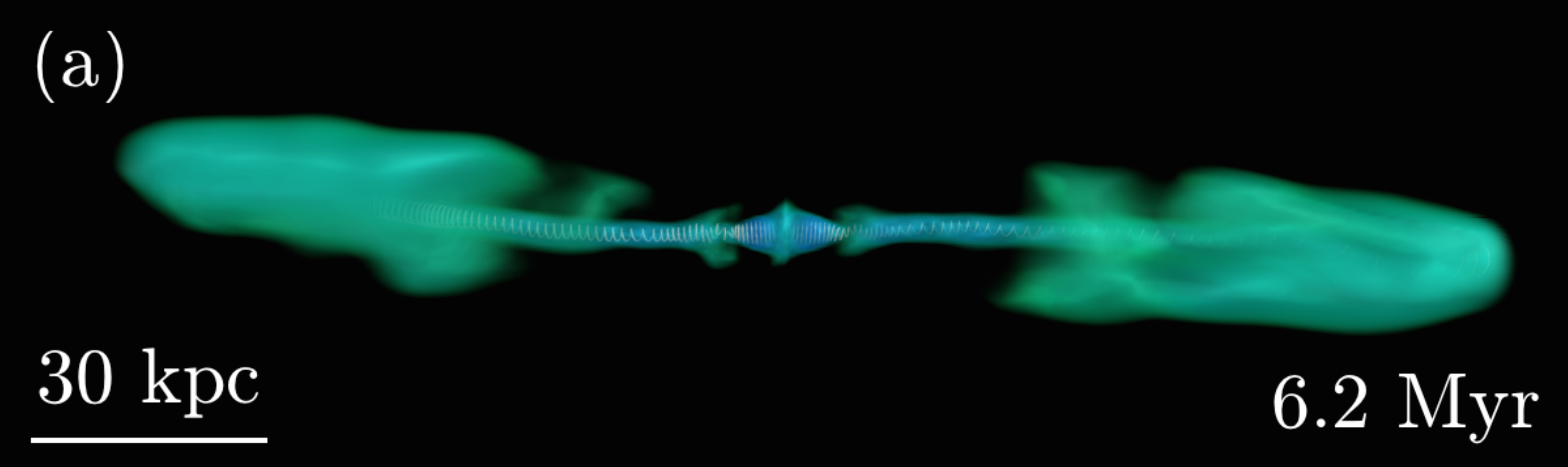}\\
\includegraphics[width=\columnwidth]{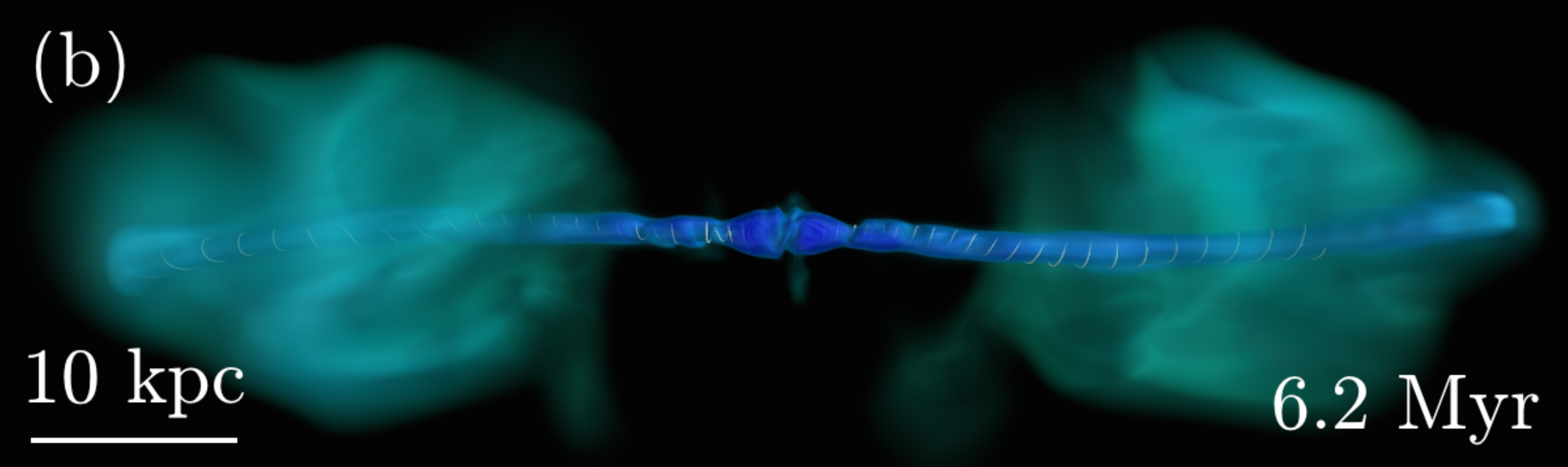}\\
\includegraphics[width=\columnwidth,trim=0 0cm 0 0cm,clip]{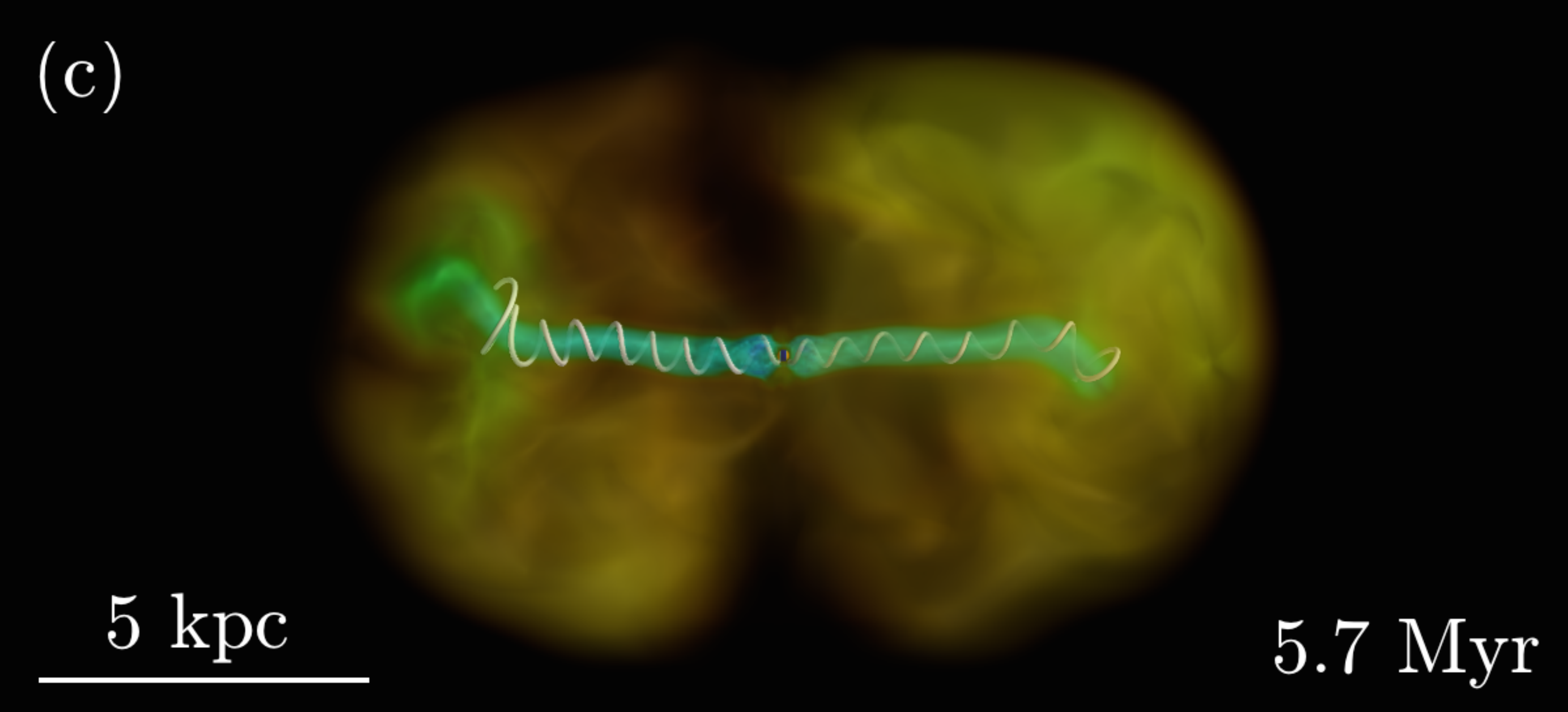}
\caption{ Volume rendering of simulated jets shows the logarithm of
  density (yellow-green shows high and blue shows low values).  White
  lines show magnetic field lines. Please see Supporting Information
  and \href{https://www.youtube.com/playlist?list=PL39mDr1uU6a7zbZldc8s92JGHPkEPHVU9}{YouTube
(link)}
  for the movies. {(panel~a):} High-power jets in model P46B2
  reach $100$~kpc distances in $6$~Myr. They are mostly straight
  apart from subtle, large-scale bends seen in FRII sources such as
  Cygnus~A; after running into the ambient medium, the jets end up
  forming strong backflows, as characteristic of FRII jet
  sources. {(panel~b):} High-power jets in model P46 reach
  $30$~kpc in $6$~Myr. Their backflows are less well-collimated
  than in model P46B2. {(panel~c):} Low-power jets in model P44,
  reach a distance of $5$~kpc in $3$~Myr. Once there, they
  succumb to a global kink instability and remain stalled at this
  distance for $3$~Myr (see Fig.~\ref{fig:zh_vs_t}). These jets
  inflate large cavities (shown in yellow) filled with a
  relativistically-hot plasma, as characteristic of FRI jet sources.}
\label{fig:dich}
\end{figure}

Figs~\ref{fig:dich}(a) and 4(b) show that the powerful jets in
models P46B2 and P46, respectively, retain their overall stability and
display the morphology characteristic of powerful FRII sources: mostly
straight, well-defined jets all the way to the points at which they
drill through the ambient medium (``hot spots'') and form powerful
backflows. Thus, so long as the jets are powerful enough, the details
of the ambient density profile do not have a very strong effect on the
jet morphology.  Note that the global kink instability causes the jets
to exhibit large-scale bends similar to those seen in some FRII
sources such as Cygnus~A \citep{1984ApJ...285L..35P}. Our simulations show that such bends
naturally develop as a result of global magnetic instabilities caused
by the interaction of the jets with the ambient medium, even without
any asymmetries imposed on the jets as they emerge from the central
engine.

The jets in the low-power model P44 show a qualitatively different
behaviour. The inset in Fig.~\ref{fig:zh_vs_t} shows a zoom-in on the
jet head dynamics in the model P44. At early times, the jets propagate
in a similar way to high-power jets: their velocity decreases as they
propagate away from the centre. However, in qualitative agreement with
the estimate~\eqref{eq:rcrit}, once the jets approach a critical
distance, $r_{\rm crit}\sim5$~kpc, a violent kink instability sets in
and stalls their propagation for a time comparable to the jet
lifetime.  The inset in Fig.~\ref{fig:zh_vs_t} shows that the stall
occurs over the time range $t=3{-}6$~Myr that is comparable to the
interval, inferred to be $\sim10$~Myr, between the subsequent
outbursts in M87 \citep{2010MNRAS.407.2046M}. Thus, at present M87
jets are likely to be stalled by the kink instability.

Fig.~\ref{fig:dich}(c) shows that the stalled low-power jets in
model P44 have a qualitatively different morphology than the
high-power jets. The 
\href{https://www.youtube.com/playlist?list=PL39mDr1uU6a7zbZldc8s92JGHPkEPHVU9}{movie}
shows the tips of low-power jets get periodically
broken off by the kink
instability. %
Such broken-off segments are potentially seen in $2$-cm VLA images of
the M87 jet at around $5$~kpc (deprojected) away from the central
BH \citep*{1989ApJ...340..698O}.  The instability-induced erratic
motion of the jet head slows down the jet propagation, spreads the jet
exhaust over a large area, and inflates quasi-spherical cavities
filled with a relativistically-hot magnetised plasma, as in
FRI sources. This is very different from powerful jets that maintain
their direction and speed and whose exhaust forms strong backflows, as
in FRII sources (Fig.~\ref{fig:dich}a,b).

The duration of the jet stall is dictated by the timescale for the
cavity to double its radius and for its jet-confining pressure to drop
substantially. After this, the jets regain their ability to propagate. However,
they still remain strongly unstable to the kink mode and develop
strong bends beyond the
critical distance, $r_{\rm crit}\sim5$~kpc. Their dynamics from then on is a mixture of
disruption and precession. Precession-induced periodic oscillations in
the position of the jet head in model P44 are especially clearly seen
in Fig.~\ref{fig:zh_vs_t} at $t=12{-}16$~Myr.
To verify the robustness of the instability, we have repeated the
low-power model P44 at twice as high effective resolution and refer to
it as model P44HR (see
Table~\ref{tab:models_grid}). The positions of jet heads in both models
agree very well (Fig.~\ref{fig:zh_vs_t}), indicating that our models are
numerically-converged. 

\vspace{-0.4cm}
\section{Discussion and Conclusions}
\label{sec:concl}
We carried out novel simulations of relativistic AGN jets propagating in an ambient medium. The simulations, for the first time, are (i)
global, (ii) 3D, (iii) magnetised, and (iv) launch jets
self-consistently via the rotation of the central object. All 4
elements are crucial for capturing the global 3D magnetic kink
instability that shapes relativistic jet feedback and interaction with
the ISM/IGM. The kink instability
grows on the timescale it takes Alfv\'en waves to
travel around the jet $\sim10$ times and becomes disruptive  when its growth time
is shorter than the fluid travel time from the
base to the tip of the jet (Sec.~\ref{sec:instab}). Thus, narrower
jets are less stable.

A jet of power $L_j$
propagating in a flat density profile
$n=n_{\rm kpc}\times(r/{\rm kpc})^{-\alpha}$, with $\alpha\sim1$
characteristic of galaxy/cluster cores, becomes progressively more collimated:
$\gamma_j\theta_j\propto r^{(\alpha-2)/6}$ (eq.~\ref{eq:lambda}).
Once $\theta_j$ drops below a critical value,
$\theta_{\rm crit}\equiv0.03/\gamma_j$ (eq.~\ref{eq:lambda}), the
instability becomes disruptive and stalls the jet. This occurs at a critical distance,
$r_{\rm crit}\propto (L_j/n_{\rm kpc})^{1/(2-\alpha)}$ (eq.~\ref{eq:rcrit}).  The instability-induced erratic motions of the jet
head slow down the jet, by spreading its exhaust over a larger area, and
inflate quasi-spherical cavities, similar to what is observed in FRI sources
(Fig.~\ref{fig:dich}c). For high-power jets, $r_{\rm crit}$ is much
larger than the flat density profile core radius, $r_{\rm core}$.
Such jets stably propagate in the flat density core region, break out of it, and remain stable to large distances.
Their exhausts form powerful backflows, as observed in FRII sources
(Fig.~\ref{fig:dich}a,b). Thus, the magnetic kink instability can
naturally lead to the \citet{fr74} dichotomy of AGN jets.

Inverting eq.~\eqref{eq:rcrit} gives us a critical jet power,
\begin{equation}
L_{j,\rm crit}=4\times10^{44}\ {\rm erg\;s^{-1}} \frac{n_{\rm kpc}}{0.2\
  \rm cm^{-3}}\left(\frac{r_{\rm core}}{30\ {\rm
      kpc}}\right)^{2-\alpha},
\label{eq:Lcrit}
\end{equation}
above which a jet will stably escape out of the flat density profile
core.
This critical power, which sets the FRI/II dividing line, increases with the
optical luminosity of a galaxy as 
$L_{j,\rm crit}\propto L_{\rm opt}^{1.8}$
\citep{ledlow_20_1996}. To verify 
whether our model reproduces this scaling, we note that
brighter galaxies have larger cores $r_{\rm
  core}\propto L_{\rm opt}^{1.1}$ \citep{1997AJ....114.1771F} and lower core densities $n_{\rm  core}\propto L_{\rm opt}^{-1/2}$ \citepalias{2014arXiv1408.3318P}.
Substituting these relations and the galaxy density profile $n_{\rm kpc}\propto n_{\rm core}({\rm kpc}/r_{\rm
  core})^{-\alpha}$ into equation~\eqref{eq:Lcrit} gives $L_{j,\rm crit}
\propto n_{\rm core} r_{\rm core}^2 \propto L_{\rm opt}^{1.7}$, in
excellent agreement with the observed scaling.

\citetalias{2014arXiv1408.3318P} performed \emph{local}
  relativistic  3D  MHD simulations of
jets in periodic boxes and found that all laterally causally-connected
jets eventually got disrupted by the magnetic kink instability.  They
argued that FRI jets are in lateral casual contact and FRII jets are
not (see also
\citealt{1995ApJS..101...29B}). However, this would imply that all our jets
beyond the recollimation point should be globally unstable, contrary to our findings. 
The difference is due to \citetalias{2014arXiv1408.3318P} adopting periodic
boundary conditions, which implied that their jets were
infinitely long ($\gg r_{\rm crit}$) and caused all
their causally-connected jets to disrupt.
\citetalias{1997MNRAS.286..215K} constructed a 2D self-similar model of
nonrelativistic hydrodynamic jets. 
They postulated phenomenologically that once shear-induced
instabilities penetrate from the shear layer
to the axis of a jet, the jet
morphology switches from FRII to FRI.  They did not verify this claim
with simulations.
The physics underlying their model is vastly different than
ours and so is their jet dynamics, however the two models share the same scaling of the jet opening angle and of the FRI/II transition with the jet
length. That these predicted scalings are so insensitive to the
underlying physical assumptions, highlights the need for
first-principles, global relativistic 3D MHD simulations to constrain the
basic jet physics. %
Here, we set the confining medium at the Bondi radius
  and neglected other jet confining effects at smaller radii, e.g.\ by
  disk wind \citep*[e.g.,][]{hk06,mck06jf,tch08}. This led to jets with order unity Lorentz factors. In
the future, we plan to include disk wind effects which may lead to 
higher values of the Lorentz factor and affect the jet stability,
morphology, and emission.
We will also explore
different density profiles, shapes of gravitational wells, and values of cluster
rotation, to identify the physical processes in sources producing
Mpc-scale FRI jets, e.g., 3C 31 \citep{hardcastle_chandra_2002}. 

We neglected gravity, buoyancy effects, and the
thermal pressure of the pristine ISM/IGM. The inclusion of the latter will only have a small effect on FRII jets, but it can
increase the confining pressure of FRI jets and decrease their stability, strengthening our
results. 
Over the simulated long time scale of
$\gtrsim15$~Myr the effects of buoyancy may eventually become important and can
expel some of the jet-inflated cavities in the form of buoyant mushroom-like
bubbles, such as those seen in the 90 cm radio image of the M87 galaxy
$\sim 20$~kpc eastward of the core \citep*{2000ApJ...543..611O}. Such slowly rising buoyant bubbles can
heat up the intracluster medium
\citep{2014Natur.515...85Z,2015MNRAS.450.4184Z,2016arXiv160502672S}. 
\vspace{-0.4cm}
\section{Acknowledgements}
\label{sec:acks}
We thank S.~Phinney, R.~Laing, S.~Markoff, A.~K\"onigl, D.~Giannios, R.~Barniol
Duran, Ashley King,
B.~McNamara, M.~Hardcastle, T.~Piran,  A.~Spitkovsky and the anonymous referee for 
discussions and helpful comments. A.T.\ was supported by NASA through Einstein Postdoctoral
Fellowship grant number PF3-140115 awarded by the Chandra X-ray Center,
operated by the Smithsonian Astrophysical Observatory for
NASA under contract NAS8-03060, and NSF through an XSEDE computational
time allocation TG-AST100040 on TACC Stampede,
Maverick, and Ranch, and NICS Darter. The
simulations presented in this work also used
the Savio cluster
provided by UCB. A.T.\ thanks Skyhouse for hospitality.  O.B.\ was supported by the Lyman Spitzer Jr.\
Fellowship, awarded by the Department of Astrophysical Sciences at
Princeton University and the Max-Planck/Princeton Center for Plasma
Physics.

\vspace{-0.4cm}
%
\vspace{-0.4cm}
\section*{Supporting Information}
Additional Supporting Information may be found in the online version
of this article and on
\href{https://www.youtube.com/playlist?list=PL39mDr1uU6a7zbZldc8s92JGHPkEPHVU9}{YouTube
(link)}:

\medskip

\noindent {\bf Movie files:} movies of Fig.~\ref{fig:dich}(a)-(c).
(\href{http://www.mnrasl.oxfordjournals.org/lookup/suppl/doi:10.1093/mnrasl/slw064/-/DC1}{http://www.mnrasl.oxford} \href{http://www.mnrasl.oxfordjournals.org/lookup/suppl/doi:10.1093/mnrasl/slw064/-/DC1}{journals.org/lookup/suppl/doi:10.1093/mnrasl/slw064/-/DC1}).

\medskip

\noindent Please note: Oxford University Press is not responsible for the
content or functionality of any supporting materials supplied by
the authors. Any queries (other than missing material) should be
directed to the corresponding author for the Letter.

\label{lastpage}
\end{document}